\documentclass[journal,12pt,onecolumn,draftclsnofoot,]{IEEEtran}
\usepackage{cite}
 \usepackage[pdftex]{graphicx}
\usepackage{mathtools}
\usepackage{amsmath}
\usepackage{amssymb}
\usepackage{caption}
\usepackage{graphicx}

\begin{document}

\title{Channel Eigenvalues and Effective Degrees of Freedom of Reconfigurable Intelligent Surfaces}

\author{Shu Sun,~\IEEEmembership{Member,~IEEE}

\thanks{Shu Sun is with the Department of Electronic Engineering, Shanghai Jiao Tong University, Shanghai, China (e-mail: ss7152@nyu.edu).}}

{}

\maketitle

\begin{abstract}
As a promising candidate technology for the next-generation wireless communications, reconfigurable intelligent surface (RIS) has gained tremendous research interest in both the academia and industry in recent years.  Only limited knowledge, however, has been obtained  about the channel eigenvlaue characteristics and degrees of freedom (DoF) of systems containing RISs. In this paper, we focus on a wireless communication system where both the transmitter and receiver are respectively equipped with an RIS. Features of eigenvalues, such as their summation and individual behavior, are investigated for both the correlation matrix of an RIS and the composite channel matrix encompassing the two RISs and the wireless channel. Furthermore, the concept of effective degrees of freedom (EDoF), i.e., the number of subchannels actively contributing to conveying information, is revisited for RIS-enabled systems. Analytical and numerical results demonstrate that the EDoF depends upon various factors including the operating SNR, physical parameters of RISs, and propagation environment. 
\end{abstract}


\IEEEpeerreviewmaketitle

\section{Introduction}
Massive MIMO (multiple-input multiple-output)\cite{Marzetta10TWC} is a pivotal physical-layer technology for the fifth-generation (5G) and beyond-5G wireless systems, which can bring immense advantages in spectral efficiency and energy efficiency \cite{Marzetta10TWC,Yan19TC,Yan20Access,Yan20TVT,Yan21IoTJ}. In typical Massive MIMO antenna arrays, the adjacent element spacing is usually half-wavelength or larger. As a natural extension of Massive MIMO, more elements may be arranged in a small form factor if the element spacing further decreases from the half-wavelength, where the whole array can be looked upon as a spatially-continuous electromagnetic aperture in its eventual form\cite{Pizzo20JSAC}. This type of extended Massive MIMO is named Holographic MIMO\cite{Pizzo20JSAC,Huang20WC}. On the other hand, reminiscent to metasurfaces in the optical field\cite{Holloway12APM,Sun_JOSA}, the sub-wavelength pattern has the potential to control incoming electromagnetic waves via anomalous reflection, refraction, polarization transformation, and other functionalities, for wireless communication purposes. Consequently, such kind of architecture is also referred to as large intelligent surface, intelligent metasurface, reconfigurable intelligent surface (RIS), among other names. In this paper, we utilize RIS as the umbrella term for all the two-dimensional (2D) sub-wavelength structures mentioned above. In order to unleash the full potentials of the RIS technology, it is necessary to understand its fundamental properties, such as the associated channel eigenvalues and spatial degrees of freedom (DoF). 

Despite the proliferation in research interests of RIS, only limited works are available in the open literature on characterizing the eigenvalue features and DoF of RIS channels. Due to the small (generally no larger than a half-wavelegnth) element spacing in an RIS, the spatial correlation is non-zero even under isotropic scattering \cite{Pizzo20JSAC,Bjornson20WCL,Sun21RISModel}. Among the early works involving antenna spatial correlation, the authors of \cite{Shiu00TC} have considered spatial correlation among multielement antennas, and derived upper and lower capacity bounds taken into account antenna correlation. Fading correlation has also been explored in \cite{Chuah02TIT} to examine the capacity growth with respect to the number of antenna elements. Nevertheless, the antenna element spacing is of half-wavelength or larger in \cite{Shiu00TC,Chuah02TIT}. The capacity of spatially dense multiple antenna systems was pioneered in \cite{Chiurtu01ITW}, which demonstrated that the capacity of such a system approches a finite limit. The spatial DoF for sufficiently dense and large RISs have been studied in \cite{Pizzo20JSAC}, and the achievable DoF for more common cases with finite element spacing and aperture areas has been investigated in \cite{Sun21RISModel}, but the influence of channel conditions, such as the SNR, on the DoF has not been considered. 

\subsection{Contributions}
In this paper, we consider wireless communications between two parallel RISs at the transmitter and receiver, respectively, and analyze the characteristics of eigenvalues of both the correlation matrix of an RIS and the composite channel matrix encompassing the two RISs and the wireless channel. The analysis herein differs from the existing literature in that a realistic, instead of assumed, spatial correlation model for RISs under isotropic scattering is employed, which has been proven in \cite{Pizzo20JSAC,Bjornson20WCL,Sun21RISModel}.  The summations, individual behavior, and distributions of the eigenvalues are investigated for a plurality of RIS sizes and element spacing. Analytical forms are provided to characterize the eigenvalue distributions. Moreover, the effective degrees of freedom (EDoF), which denotes the number of subchannels actively conveying information \cite{Shiu00TC}, is studied and compared with the theoretical asymptotic DoF for sufficiently large and dense RISs without considering SNR conditions, in terms of their values and effect on channel capacity. Numerical results demonstrate that the EDoF relies on the operating SNR, in addition to spatial correlation which is in turn jointly determined by the physical parameters of RISs and propagation environment. The contributions in this paper are useful to channel estimation and beamforming design for RISs \cite{Sun21RIS}.

\subsection{Outline of the Paper and Notation}
The remainder of this paper is organized as follows. In Section II, we describe the system model containing RISs and identify the terms affecting the channel capacity. Asymptotic analysis of eigenvalues for both the correlation matrix at the RIS and the equivalent composite channel matrix is carried out in Section III. In Section IV, we systematically investigate the EDoF under various settings. Conclusions are drawn in Section V. 

The following notations will be utilized throughout the paper: $\textbf{A}$ for matrix, $\textbf{a}$ for column vector, $\textbf{a}_k$ for the $k$th column of $\textbf{A}$, $[\textbf{A}]_{i,j}$ for the $(i,j)$th entry of $\textbf{A}$, $\textbf{a}_k(i)$ for the $i$th entry of $\textbf{a}_k$, $\textbf{A}^\text{H}$ for transpose conjugate of $\textbf{A}$, $\det(\textbf{A})$ for determinant of the square matrix $\textbf{A}$, $\text{tr}(\textbf{A})$ for the trace of $\textbf{A}$, and $\textbf{I}_N$ for the $N\times N$ identity matrix. 

\section{System Model}
\begin{figure}
	\centering
	\includegraphics[width=0.7\columnwidth]{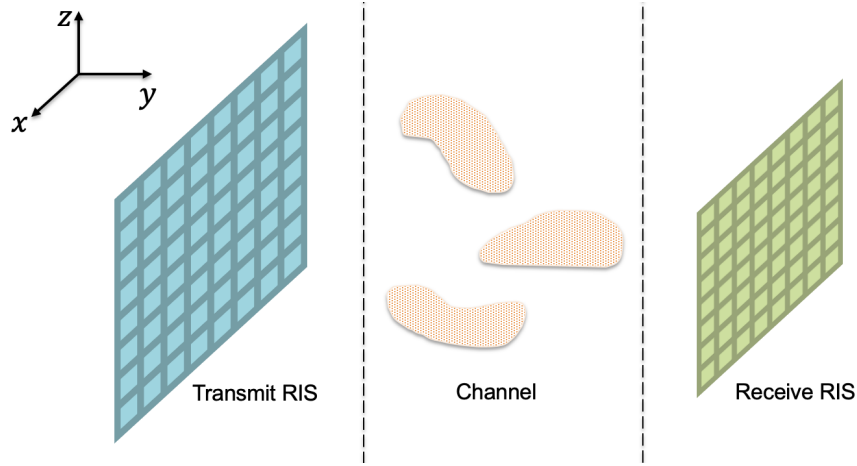}
	\caption{Orientations of the transmit and receive RISs with respect to the associated coordinate system.}
	\label{fig:fig1}	
\end{figure}
We consider a point-to-point wireless communication system where the transmitter and receiver are each equipped with an RIS with $N_\text{T}$ and $N_\text{R}$ elements, respectively, where the two RISs are parallel with each other, as illustrated in Fig.~\ref{fig:fig1}. The horizontal and vertical lengths of the RISs are $L_{\text{T},x},L_{\text{T},z}$ and $L_{\text{R},x},L_{\text{R},z}$, respectively. For any subband of the radio bandwidth within which the channel response can be considered frequency-flat, the discrete-time system model is given by
\begin{equation}\label{eq:y}
\textbf{y}=\sqrt{\rho}\textbf{H}\textbf{x}+\textbf{n}
\end{equation}

\noindent where $\textbf{y}\in\mathbb{C}^{N_\text{R}\times1}$, $\textbf{H}\in\mathbb{C}^{N_\text{R}\times N_\text{T}}$, and $\textbf{x}\in\mathbb{C}^{N_\text{T}\times1}$ denote the received signal vector, channel matrix, and transmitted signal vector, respectively. $\rho$ represents the signal-to-noise ratio (SNR) at the receiver that entails the large-scale fading coefficient. Besides, $\textbf{n}\in\mathbb{C}^{N_\text{R}\times1}$ represents the noise vector whose entries are independently and identically distributed (i.i.d.) circularly symmetric complex Gaussian random variables with mean zero. Owing to the compact arrangement of the elements in the RISs at both link ends, and the fact that the spatial correlation at each RIS is determined by the immediate surroundings to the RIS and not affected by the spatial correlation at the other end of the link in the far-field region \cite{Moustakas00Science,Chuah02TIT,Tulino05TIT}, the spatial correlation matrices at the transmitter and receiver are separable. Therefore, $\textbf{H}$ can be factorized as\cite{Tulino05TIT,Wei05TWC}
\begin{equation}\label{eq:H1}
\textbf{H}=\textbf{R}_{\text{R}}^{1/2}\textbf{H}_w\textbf{R}_{\text{T}}^{1/2}
\end{equation}

\noindent in which $\textbf{R}_{\text{R}}\in\mathbb{C}^{N_\text{R}\times N_\text{R}}$ and $\textbf{R}_{\text{T}}\in\mathbb{C}^{N_\text{T}\times N_\text{T}}$ are the correlation matrices at the receive and transmit RISs, respectively, while $\textbf{H}_w\in\mathbb{C}^{N_\text{R}\times N_\text{T}}$ is i.i.d. Rayleigh-faded under isotropic scattering with the subscript $w$ standing for "white". Denote the covariance matrix of the input signal vector $\textbf{x}$ as $\textbf{Q}=\mathbb{E}\{\textbf{x}\textbf{x}^\text{H}\}$, the ergodic capacity of  (\ref{eq:y}) in bit/s/Hz is \cite{Telatar99ETT}
\begin{equation}\label{eq:C1}
C=\max_{\textbf{Q}:\text{tr}(\textbf{Q})\leq1}\mathbb{E}\Big\{\log_2\det\big(\textbf{I}_{N_\text{R}}+\rho\textbf{H}\textbf{Q}\textbf{H}^\text{H}\big)\Big\}.
\end{equation}

\noindent With linear operations adopted at both the transmitter and receiver, (\ref{eq:y}) can be converted into an equivalent system involving $N_\text{S}$ decoupled single-input single-output (SISO) subchannels, where each subchannel corresponds to a spatial eigenmode \cite{Shiu00TC}, and $N_\text{S}$ satisfies 
\begin{equation}\label{eq:Ns}
N_\text{S}\leq\min(N_\text{T},N_\text{R}).
\end{equation}

\noindent If assuming instantaneous channel state information (CSI) available at the receiver while no channel matrix information but just large-scale path loss available at the transmitter, which is one of the common scenarios in practice \cite{Shiu00TC}, the ergodic capacity in (\ref{eq:C1}) is achieved by setting $\textbf{Q}$ as $\frac{1}{N_\text{S}}\textbf{I}_{N_\text{T}}$, implying allocating the transmit power equally among the subchannels. Therefore, (\ref{eq:C1}) is transformed to
\begin{equation}\label{eq:C2}
C=\mathbb{E}\Big\{\log_2\det\big(\textbf{I}_{N_\text{R}}+\frac{\rho}{N_\text{S}}\textbf{H}\textbf{H}^\text{H}\big)\Big\}
\end{equation}

\noindent It is worth noting that the uniform power allocation strategy does not require the knowledge of $\textbf{H}$, rendering it appealing in a variety of systems where it is inconvenient or impossible to acquire the channel matrix.  Plugging (\ref{eq:H1}) into (\ref{eq:C2}), we obtain
\begin{equation}\label{eq:C3}
\begin{split}
C=&~\mathbb{E}\Big\{\log_2\det\Big(\textbf{I}_{N_\text{R}}+\frac{\rho}{N_\text{S}}\textbf{R}_{\text{R}}^{1/2}\textbf{H}_w\textbf{R}_{\text{T}}^{1/2}\big(\textbf{R}_{\text{R}}^{1/2}\textbf{H}_w\textbf{R}_{\text{T}}^{1/2}\big)^\text{H}\Big)\Big\}\\
=&~\mathbb{E}\Big\{\log_2\det\Big(\textbf{I}_{N_\text{R}}+\frac{\rho}{N_\text{S}}\textbf{R}_{\text{R}}\textbf{H}_w\textbf{R}_{\text{T}}\textbf{H}_w^\text{H}\Big)\Big\}\\
\end{split}
\end{equation}

\noindent Since $\textbf{H}_w$ is isotropic, the distributions of $\textbf{U}\textbf{H}_w$ and $\textbf{H}_w\textbf{U}$ are identical to $\textbf{H}_w$ for any deterministic unitary matrix $\textbf{U}$. Given that $\textbf{R}_{\text{T}}$ and $\textbf{R}_{\text{R}}$ are Hermitian, they can be decomposed as
\begin{equation}\label{eq:RR}
\begin{split}
\textbf{R}_{\text{T}}=\textbf{U}\textbf{D}_{\text{T}}\textbf{U}^\text{H},~\textbf{R}_{\text{R}}=\textbf{V}\textbf{D}_{\text{R}}\textbf{V}^\text{H}
\end{split}
\end{equation}

\noindent where $\textbf{U}\in\mathbb{C}^{N_\text{T}\times N_\text{T}}$ and $\textbf{V}\in\mathbb{C}^{N_\text{R}\times N_\text{R}}$ are unitary, while $\textbf{D}_{\text{T}}\in\mathbb{C}^{N_\text{T}\times N_\text{T}}$ and $\textbf{D}_{\text{R}}\in\mathbb{C}^{N_\text{R}\times N_\text{R}}$ are diagonal matrices composed of the eigenvalues of $\textbf{R}_{\text{T}}$ and $\textbf{R}_{\text{R}}$, respectively, in non-increasing order. Combining (\ref{eq:C3}) and (\ref{eq:RR}) yields
\begin{equation}\label{eq:C4}
\begin{split}
C=&~\mathbb{E}\Big\{\log_2\det\Big(\textbf{I}_{N_\text{R}}+\frac{\rho}{N_\text{S}}\textbf{V}\textbf{D}_{\text{R}}\textbf{V}^\text{H}\textbf{H}_w\textbf{U}\textbf{D}_{\text{T}}\textbf{U}^\text{H}\textbf{H}_w^\text{H}\Big)\Big\}\\
=&~\mathbb{E}\Big\{\log_2\det\Big(\textbf{I}_{N_\text{R}}+\frac{\rho}{N_\text{S}}\textbf{D}_{\text{R}}\textbf{V}^\text{H}\textbf{H}_w\textbf{U}\textbf{D}_{\text{T}}\textbf{U}^\text{H}\textbf{H}_w^\text{H}\textbf{V}\Big)\Big\}\\
\stackrel{\mathcal{D}}{=}&~\mathbb{E}\Big\{\log_2\det\Big(\textbf{I}_{N_\text{R}}+\frac{\rho}{N_\text{S}}\textbf{D}_{\text{R}}\textbf{H}_w\textbf{D}_{\text{T}}\textbf{H}_w^\text{H}\Big)\Big\}\\
\end{split}
\end{equation}

\noindent with $\mathcal{D}$ meaning "in distribution". The third equation is by virtue of the fact that the entries of $\textbf{H}_w$ are i.i.d. complex Gaussian random variables, whose joint distribution will not change when $\textbf{H}_w$ is multiplied by a unitary matrix. 

It is evident from (\ref{eq:C4}) that the channel capacity is dependent on the eigenvalues of the random matrix $\textbf{D}_{\text{R}}\textbf{H}_w\textbf{D}_{\text{T}}\textbf{H}_w^\text{H}$ and an appropriate selection of $N_\text{S}$ which are the core focus of this article. In the next section, we will first investigate the characteristics of the eigenvalues of $\textbf{R}_{\text{T}}$ and $\textbf{R}_{\text{R}}$ at the RISs, followed by the analysis on the asymptotic behavior of the eigenvalues of $\textbf{D}_{\text{R}}\textbf{H}_w\textbf{D}_{\text{T}}\textbf{H}_w^\text{H}$ and thus the capacity. 

\section{Asymptotic Analysis of Eigenvalues}
\subsection{Characteristics of Eigenvalues of Correlation Matrix at the RIS}\label{sec:evR}
Substituting $\textbf{R}_{\text{T}}$ and $\textbf{R}_{\text{R}}$ in (\ref{eq:H1}) with (\ref{eq:RR}) results in
\begin{equation}\label{eq:H2}
\begin{split}
\textbf{H}=\textbf{V}\textbf{D}_{\text{R}}^{1/2}\textbf{V}^\text{H}\textbf{H}_w\textbf{U}\textbf{D}_{\text{T}}^{1/2}\textbf{U}^\text{H}=\textbf{V}\textbf{D}_{\text{R}}^{1/2}\tilde{\textbf{H}}_w\textbf{D}_{\text{T}}^{1/2}\textbf{U}^\text{H}\\
\end{split}
\end{equation}

\noindent where $\tilde{\textbf{H}}_w=\textbf{V}^\text{H}\textbf{H}_w\textbf{U}\in\mathbb{C}^{N_\text{R}\times N_\text{T}}$. Furthermore, denoting $\textbf{D}_{\text{R}}^{1/2}\tilde{\textbf{H}}_w\textbf{D}_{\text{T}}^{1/2}$ as $\tilde{\textbf{H}}$, we obtain
\begin{equation}\label{eq:H3}
\begin{split}
\textbf{H}=\textbf{V}\tilde{\textbf{H}}\textbf{U}^\text{H}
\end{split}
\end{equation}

\noindent Employing the unitary property of $\textbf{V}$ and $\textbf{U}$, (\ref{eq:H3}) is tantamount to 
\begin{equation}\label{eq:H4}
\begin{split}
\tilde{\textbf{H}}=\textbf{V}^\text{H}\textbf{H}\textbf{U}
\end{split}
\end{equation}

\noindent which can be recast as
\begin{equation}\label{eq:H5}
\begin{split}
[\tilde{\textbf{H}}]_{k,l}=\sum_{i=1}^{N_\text{R}}\sum_{j=1}^{N_\text{T}}\textbf{v}^\text{H}_k(i)[\textbf{H}]_{i,j}\textbf{u}_l(j)
\end{split}
\end{equation}

\noindent Since $\textbf{V}$ and $\textbf{U}$ are deterministic unitary matrices containing a set of complete orthonormal discrete basis functions formed by the eigenfunctions of $\textbf{R}_{\text{R}}$ and $\textbf{R}_{\text{T}}$, respectively, $\tilde{\textbf{H}}$ can be regarded as the Karhunen-Loeve transform (KLT) of $\textbf{H}$ \cite{Tulino05TIT}. $\{\textbf{v}_k\}$ and $\{\textbf{u}_l\}$ are the KLT kernel satisfying
\begin{equation}\label{eq:EVD1}
\begin{split}
\sum_{i'=1}^{N_\text{R}}[\textbf{R}_{\text{R}}]_{i,i'}\textbf{v}_k(i')=\alpha_k\textbf{v}_k(i),~\sum_{j'=1}^{N_\text{T}}[\textbf{R}_{\text{T}}]_{j,j'}\textbf{u}_l(j')=\beta_l\textbf{u}_l(j)
\end{split}
\end{equation}

\noindent where $\alpha_k$ and $\beta_l$ denote the $k$th eigenvalue of $\textbf{R}_{\text{R}}$ and $l$th eigenvalue of $\textbf{R}_{\text{T}}$, respectively. 

To investigate the features of the eigenvalues of $\textbf{R}_{\text{R}}$ and $\textbf{R}_{\text{T}}$, let's focus on the transmitter side, since the receiver side can be treated similarly. For ease of exposition, we shall drop the subscript $\text{T}$ by the following notations: $\textbf{R}=\textbf{R}_{\text{T}},N=N_\text{T},L_x=L_{\text{T},x}$, and $L_z=L_{\text{T},z}$. Without loss of generality, let $g(r)$, which maps $r\in[-\sqrt{L_x^2+L_z^2},\sqrt{L_x^2+L_z^2}]$ to the real line, be the normalized (i.e., $g(0)=1$) spatial correlation function for an RIS of fixed dimension, such that
\begin{equation}\label{eq:R2}
\begin{split}
g\bigg(\frac{i-j}{N-1}\sqrt{L_x^2+L_z^2}\bigg)=\frac{[\textbf{R}]_{i,j}}{N},~i,j=1,...,N
\end{split}
\end{equation}

\noindent Let $\alpha_k^{\textbf{A}}$ represent the $k$th largest eigenvalue of the matrix $\textbf{A}$. Due to the normalization $g(0)=1$, we have 
\begin{equation}\label{eq:R3}
\begin{split}
\sum_{k=1}^{N}\alpha_k^{\textbf{R}/N}=\text{tr}\bigg(\frac{\textbf{R}}{N}\bigg)=1,~\forall N
\end{split}
\end{equation}

\noindent In addition, $\{\alpha_k^{(N)}\}$ will be converging to the point spectrum (i.e., eigenvalues) $\{\alpha_k^{(\infty)}\}$ of the non-negative definite Hermitian operator $\tilde{g}(x,y)=g\big((x-y)\sqrt{L_x^2+L_z^2}\big)$ where $x,y\in[0,1]$ \cite{Chiurtu01ITW,Wei05TWC}. $\{\alpha_k^{(\infty)}\}$ can be determined by
\begin{equation}\label{eq:R4}
\begin{split}
\int_{0}^{1}g\big((x-y)\sqrt{L_x^2+L_z^2}\big)\phi_k(y)dy=\alpha_k^{(\infty)}\phi_k(x),\\
x\in[0,1],k=0,1,...,\infty
\end{split}
\end{equation}

\noindent in which $\{\phi_k(x)\}$ are the eigenfunctions of the operator $\tilde{g}(x,y)$. The nonzero eigenvalues of $\tilde{g}(x,y)$ have finite multiplicity and form a sequence approaching zero if they are denumerable infinite in number \cite{Riesz55,Wei05TWC}. Denote the number of nonzero eigenvalues of $\textbf{R}/N$ as $f(N)$, then $f(N)\sim o(N)$ \cite{Wei05TWC}. Additionally, the speed of $f(N)/N\to0$ relies on the smoothness of $g(r)$ in the sense of continuous differentiability of various orders, and smoother $g(r)$ gives rise to a more rapid convergence of $f(N)/N\to0$. 

\begin{figure}
	\centering
	\includegraphics[width=0.7\columnwidth]{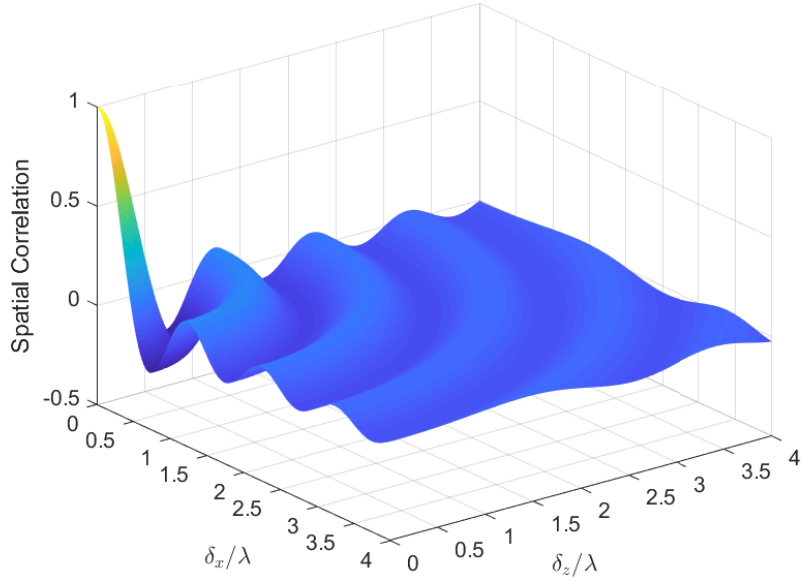}
	\caption{Spatial correlation among RIS elements under isotropic scattering, where $\lambda$ denotes the wavelength, $\delta_x/\lambda$ and $\delta_z/\lambda$ represent the element spacing along the $x$ and $z$ directions, respectively.}
	\label{fig:fig2}	
\end{figure}

It has been proven in \cite{Bjornson20WCL,Sun21RISModel} that the spatial correlation matrix at an RIS under isotropic scattering can be characterized as a sinc function as follows
\begin{equation}\label{eq:R1}
\begin{split}
[\textbf{R}]_{m,n}=\text{sinc}\bigg(\frac{2||\textbf{d}_m-\textbf{d}_n||}{\lambda}\bigg),~m,n=1,...,N
\end{split}
\end{equation}

\noindent where $\text{sinc}(x)=\frac{\text{sin}(\pi x)}{\pi x}$ is the sinc function, $\textbf{d}_m$ and $\textbf{d}_n$ denote the coordinates of the $m$-th and $n$-th RIS element, respectively. The behavior of $\textbf{R}$ is depicted in Fig.~\ref{fig:fig2} for element spacing up to four times the wavelength $\lambda$. It is observed from (\ref{eq:R1}) and Fig.~\ref{fig:fig2} that the spatial correlation is minimal only for some element spacing, instead of between any two elements, thus the i.i.d. Rayleigh fading model is not applicable in such a system \cite{Pizzo20JSAC,Bjornson20WCL,Sun21RISModel}. 
\begin{figure}
	\centering
	\includegraphics[width=0.7\columnwidth]{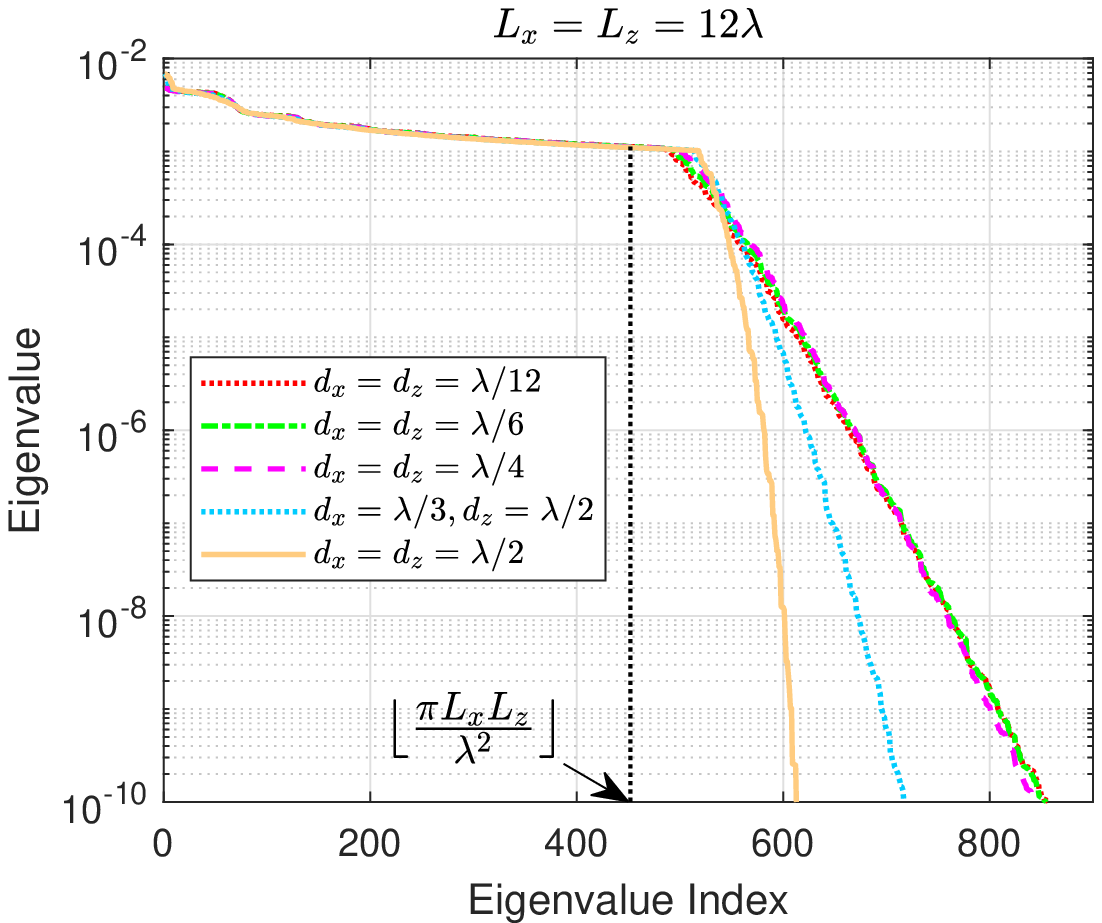}
	\caption{Eigenvalues of $\textbf{R}/N$ in non-increasing order for various element spacing $d_x$ and $d_z$ with $L_x=L_z=12\lambda$. Also depicted is the asymptotic spatial degrees of freedom (DoF) $\lfloor\frac{\pi L_xL_z}{\lambda^2}\rfloor$ derived in \cite{Pizzo20SPAWC} for $\text{min}(L_x,L_z)/\lambda\to\infty$.}
	\label{fig:fig3}	
\end{figure}

Fig.~\ref{fig:fig3} illustrates the eigenvalues of $\textbf{R}/N$ in non-increasing order for various $N$, or equivalently element spacing, with $L_x=L_z=12\lambda$. Besides, the dotted vertical line represents the asymptotic spatial DoF $\lfloor\frac{\pi L_xL_z}{\lambda^2}\rfloor$ derived in \cite{Pizzo20SPAWC} for $\text{min}(L_x,L_z)/\lambda\to\infty$. To provide a more quantitative examination and comparison of the eigenvalues for different element spacing, some selected eigenvalues are listed in Table~\ref{tbl:evR}. A few key remarks can be drawn from Fig.~\ref{fig:fig3} and Table~\ref{tbl:evR}: First, the i.i.d. Rayleigh fading channel has almost identical non-trivial eigenvalues whose amount equals the number of antenna elements deployed, while the correlated channel has uneven and fewer dominant eigenvalues and smaller rank. Second, as mentioned above, all the eigenvalues of $\textbf{R}/N$ converge for relatively large $N$, e.g., roughly starting from $d_x=d_z=\lambda/6$. Moreover, the dominant eigenvalues (whose indices are within around $\lfloor\frac{\pi L_xL_z}{\lambda^2}\rfloor$) converge even for small $N$ corresponding to half-wavelength spacing. Additionally, as demonstrated in detail in \cite{Sun21RISModel}, the asymptotic spatial DoF $\lfloor\frac{\pi L_xL_z}{\lambda^2}\rfloor$ has non-negligible approximation errors for limited RIS aperture sizes which are often encountered in practice. 
\begin{table*}
		\caption{Selected eigenvalues of $\textbf{R}/N$ in non-increasing order for various element spacing $d_x$ and $d_z$ with $L_x=L_z=12\lambda$}
		\label{tbl:evR}
		\centering
\begin{center}
	\begin{tabular}{|p{0.03\textwidth}||p{0.13\textwidth}|p{0.13\textwidth}|p{0.13\textwidth}|p{0.13\textwidth}|p{0.13\textwidth}|}
		\hline
		 & \multicolumn{5}{c|}{$\alpha_k$}\\ 
		\hline
		$k$ & $d_x=d_z=\frac{\lambda}{2}$ & $d_x=\frac{\lambda}{3},d_z=\frac{\lambda}{2}$ & $d_x=d_z=\frac{\lambda}{4}$ & $d_x=d_z=\frac{\lambda}{6}$ & $d_x=d_z=\frac{\lambda}{12}$ \\ 
		\hline\hline
		$1$ & $0.00679$ &$0.00688$&$0.00479$&$0.00484$&$0.00489$ \\ 
		\hline
		$2$ & $0.00679$ &$0.00663$&$0.00479$&$0.00484$&$0.00489$\\ 
		\hline
		$3$ & $0.00655$ &$0.00620$&$0.00479$&$0.00483$&$0.00488$ \\ 
		\hline
		$4$ & $0.00655$ &$0.00557$&$0.00479$&$0.00483$&$0.00488$ \\ 
		\hline
		$5$ & $0.00614$ &$0.00479$&$0.00457$&$0.00461$&$0.00466$\\ 
		\hline
		$6$ & $0.00614$ &$0.00471$&$0.00457$&$0.00461$&$0.00466$ \\ 
		\hline
		$7$ & $0.00554$ &$0.00471$&$0.00457$&$0.00461$&$0.00466$ \\ 
		\hline
		$8$ & $0.00554$ &$0.00469$&$0.00457$&$0.00461$&$0.00466$ \\ 
		\hline
		$9$ & $0.00479$ &$0.00469$&$0.00450$&$0.00454$&$0.00459$ \\ 
		\hline
		$10$ & $0.00478$ &$0.00466$&$0.00450$&$0.00454$&$0.00458$\\ 
		\hline
		$50$ & $0.00380$ &$0.00393$&$0.00412$&$0.00415$&$0.00418$ \\ 
		\hline
		$100$ & $0.00245$ &$0.00242$&$0.00242$&$0.00244$&$0.00246$ \\ 
		\hline
		$200$ & $0.00169$ &$0.00171$&$0.00173$&$0.00172$&$0.00172$\\ 
		\hline
		$300$ & $0.00137$ &$0.00138$&$0.00141$&$0.00142$&$0.00143$\\ 
		\hline
		$400$ & $0.00117$ &$0.00118$&$0.00119$&$0.00120$&$0.00121$\\ 
		\hline
		$500$ & $0.00104$ &$0.00105$&$0.00099$&$0.00087$&$0.00075$ \\ 
		\hline
		$600$ & $1.24\times10^{-8}$ &$6.82\times10^{-6}$&$2.23\times10^{-5}$&$1.93\times10^{-5}$&$1.56\times10^{-5}$ \\ 
		\hline
		$700$ & N/A &$6.05\times10^{-10}$&$1.82\times10^{-7}$&$2.12\times10^{-7}$&$1.87\times10^{-7}$ \\ 
		\hline
		$800$ & N/A &$2.97\times10^{-15}$&$1.08\times10^{-9}$&$1.66\times10^{-9}$&$1.64\times10^{-9}$  \\ 
		\hline
		$900$ & N/A &$0$&$4.16\times10^{-12}$&$9.88\times10^{-12}$&$1.13\times10^{-11}$ \\ 
		\hline
		$1000$ & N/A &N/A&$1.40\times10^{-14}$&$5.19\times10^{-14}$&$7.47\times10^{-14}$\\ 
		\hline
		$2000$ & N/A &N/A&$0$&$1.05\times10^{-18}$&$1.41\times10^{-18}$\\ 
		\hline
		$3000$ & N/A &N/A&N/A&$2.11\times10^{-19}$&$1.08\times10^{-18}$\\ 
		\hline
	\end{tabular}
\end{center}
\end{table*}

\subsection{Characteristics of Eigenvalues of Channel Matrix}
In this section, we study the eigenvalue behavior of the channel matrix $\textbf{D}_{\text{R}}\textbf{H}_w\textbf{D}_{\text{T}}\textbf{H}_w^\text{H}$ in (\ref{eq:C4}). Define the matrix $\tilde{\textbf{B}}\in\mathbb{C}^{N_\text{R}\times N_\text{R}}$ as the equivalent channel matrix normalized by the number of RIS elements at both the transmitter and receiver, i.e. 
\begin{equation}\label{eq:B1}
\begin{split}
\tilde{\textbf{B}}=\frac{\textbf{D}_{\text{R}}\textbf{H}_w\textbf{D}_{\text{T}}\textbf{H}_w^\text{H}}{N_\text{T}N_\text{R}}=\tilde{\textbf{D}}_{\text{R}}\textbf{H}_w\tilde{\textbf{D}}_{\text{T}}\textbf{H}_w^\text{H}.
\end{split}
\end{equation}

\noindent where $\tilde{\textbf{D}}_{\text{R}}=\textbf{D}_{\text{R}}/N_\text{R}$ and $\tilde{\textbf{D}}_{\text{T}}=\textbf{D}_{\text{T}}/N_\text{T}$. It is relevant to first investigate the summation of the eigenvalues $\big\{\alpha_k^{\tilde{\textbf{B}}}\big\}$ of $\tilde{\textbf{B}}$. Note that 
\begin{equation}\label{eq:B3}
\begin{split}
\sum_{k=1}^{N_\text{R}}\alpha_k^{\tilde{\textbf{B}}}=\text{tr}\big(\tilde{\textbf{B}}\big)=\sum_{i=1}^{N_\text{R}}\big[\tilde{\textbf{B}}\big]_{i,i}
\end{split}
\end{equation}

\noindent where $\big[\tilde{\textbf{B}}\big]_{i,i}$ is calculated as
\begin{equation}\label{eq:B4}
\begin{split}
\big[\tilde{\textbf{B}}\big]_{i,i}=&~\big[\tilde{\textbf{D}}_\text{R}\big]_{i,i}\sum_{l=1}^{N_\text{T}}\big|[\textbf{H}_w]_{i,l}\big|^2\big[\tilde{\textbf{D}}_\text{T}\big]_{l,l}\\
\stackrel{(a)}{\approx}&~\big[\tilde{\textbf{D}}_\text{R}\big]_{i,i}\sum_{l=1}^{N_\text{T}}\big[\tilde{\textbf{D}}_\text{T}\big]_{l,l}\\
\stackrel{(b)}{=}&~\big[\tilde{\textbf{D}}_\text{R}\big]_{i,i}
\end{split}
\end{equation}

\noindent where $(a)$ is because the variance of each element in $\textbf{H}_w$ is 1, and $(b)$ is ascribed to the normalized eigenvalue summation of $\textbf{R}_\text{T}/N_\text{T}$ as shown in (\ref{eq:R3}). Plugging (\ref{eq:B4}) into (\ref{eq:B3}) gives
\begin{equation}\label{eq:B5}
\begin{split}
\sum_{k=1}^{N_\text{R}}\alpha_k^{\tilde{\textbf{B}}}\approx\sum_{i=1}^{N_\text{R}}\big[\tilde{\textbf{D}}_\text{R}\big]_{i,i}=1
\end{split}
\end{equation}

\noindent which proves that the summation of the eigenvalues of $\tilde{\textbf{B}}$ is approximately 1 which equals the respective eigenvalue summation of $\textbf{R}_\text{T}/N_\text{T}$ and $\textbf{R}_\text{R}/N_\text{R}$. The cumulative distribution function (CDF) $F(\alpha^{\tilde{\textbf{B}}})$ of an arbitrary unordered eigenvalue of $\tilde{\textbf{B}}$ can be derived from the result in \cite[(15)]{Simon04PRE}, which is expressed as
\begin{equation}\label{eq:CDF1}
\begin{split}
F\Big(\alpha^{\tilde{\textbf{B}}}\Big)=\frac{1}{N_\text{R}}-\frac{Q_0\Big(\alpha^{\tilde{\textbf{B}}}\Big)R_0}{N_\text{T}N_\text{R}}\sum_{n=1}^{N_\text{R}}\det\bigg(\Big[\textbf{K}^{(n)}\Big]_{i,j}\bigg)
\end{split}
\end{equation}

\noindent where
\begin{equation}\label{eq:Q}
\begin{split}
Q_v(z)^{-1}=&~\Delta_{N_\text{R}}\Big(\alpha^{\tilde{\textbf{D}}_{\text{R}}}\Big)\Delta_{N_\text{T}}\Big(\alpha^{\tilde{\textbf{D}}_{\text{T}}}\Big)(-z)^{N_\text{T}(N_\text{T}-1)/2}J_v
\end{split}
\end{equation}

\noindent in which $\Delta$ represents a V-dimensional Vandermonde determinant given by
\begin{equation}\label{eq:Delta}
\begin{split}
\Delta_V(x)=\prod_{1\leq i<j\leq V}(x_j-x_i)=\det\big[x_j^{i-1}\big]
\end{split}
\end{equation}

\noindent and $J_v$ is calculated as
\begin{equation}\label{eq:J}
\begin{split}
J_v=\prod_{i=1}^{N_\text{R}-1}(v+i)^i
\end{split}
\end{equation}

\noindent $R_v$ in (\ref{eq:CDF1}) is defined as
\begin{equation}\label{eq:R}
R_v =  \left\{ \begin{array}{rcl}
&\prod\limits_{j=1}^{N_\text{R}-N_\text{T}-1}(N_\text{R}+v-j)^{N_\text{R}-N_\text{T}-j},&N_\text{R}>N_\text{T}+1 \\ &1,&N_\text{R}\leq N_\text{T}+1
\end{array}\right.
\end{equation}

\noindent The $N_\text{R}\times N_\text{R}$ matrix $\textbf{K}^{(n)}$ in (\ref{eq:CDF1}) is given by
\begin{equation}\label{eq:K}
\Big[\textbf{K}^{(n)}\Big]_{i,j}=\left\{ \begin{array}{rcl}
&g\Big(\alpha_i^{\tilde{\textbf{D}}_{\text{R}}}\alpha_j^{\tilde{\textbf{D}}_{\text{T}}};N_\text{R},\alpha^{\tilde{\textbf{B}}}\Big),&n\neq i \\ &(N_\text{R}-1)!e^{-\alpha_i^{\tilde{\textbf{D}}_{\text{R}}}\alpha_j^{\tilde{\textbf{D}}_{\text{T}}}\alpha^{\tilde{\textbf{B}}}},&n=i
\end{array}\right.
\end{equation}

\noindent in which for integer $M>0$, the function $g$ is defined as \cite[(13)]{Simon04PRE}
\begin{equation}\label{eq:g}
\begin{split}
g(x;M,z)=x^{N_\text{R}-M}(M-1)!\sum_{m=0}^{M-1}\frac{(-zx)^m}{m!}
\end{split}
\end{equation}

Next, we explore the behavior of individual eigenvalues of $\tilde{\textbf{B}}$ in (\ref{eq:B1}), and start with the matrix without the term $\textbf{D}_{\text{R}}$ in $\tilde{\textbf{B}}$, i.e. 
\begin{equation}\label{eq:F1}
\begin{split}
\textbf{F}=\frac{\textbf{H}_w\textbf{D}_{\text{T}}\textbf{H}_w^\text{H}}{N_\text{T}N_\text{R}}=\frac{1}{N_\text{R}}\textbf{H}_w\tilde{\textbf{D}}_{\text{T}}\textbf{H}_w^\text{H}
\end{split}
\end{equation}

\noindent which yields
\begin{equation}\label{eq:F2}
\begin{split}
\alpha_k^{\textbf{F}}=\alpha_k^{\frac{1}{N_\text{R}}\textbf{H}_w\tilde{\textbf{D}}_{\text{T}}\textbf{H}_w^\text{H}}=\alpha_k^{\frac{1}{N_\text{R}}\textbf{H}_w^\text{H}\textbf{H}_w\tilde{\textbf{D}}_{\text{T}}},~k=1,...,\text{rank}(\textbf{F})
\end{split}
\end{equation}

\noindent where the second equality stems from the fact that the (nonzero) eigenvalues of the matrix $\textbf{A}\textbf{B}$ equal those of $\textbf{B}\textbf{A}$ (with qualifying dimensions of $\textbf{A}$ and $\textbf{B}$). It has been proved in \cite{Bai99AP} that for a matrix with the form $\frac{1}{N_\text{R}}\textbf{H}_w^\text{H}\textbf{H}_w\tilde{\textbf{D}}_{\text{T}}$ in (\ref{eq:F2}), the following inequality holds:
\begin{equation}\label{eq:ev2}
\begin{split}
\alpha_k^{\tilde{\textbf{D}}_{\text{T}}}\alpha_{\text{rank}(\textbf{F})}^{\frac{1}{N_\text{R}}\textbf{H}_w^\text{H}\textbf{H}_w}\leq\alpha_k^{\textbf{F}}\leq\alpha_k^{\tilde{\textbf{D}}_{\text{T}}}\alpha_1^{\frac{1}{N_\text{R}}\textbf{H}_w^\text{H}\textbf{H}_w},~k=1,...,\text{rank}(\textbf{F})
\end{split}
\end{equation}

\noindent where $\alpha_{\text{rank}(\textbf{F})}^{\frac{1}{N_\text{R}}\textbf{H}_w^\text{H}\textbf{H}_w}$ and $\alpha_1^{\frac{1}{N_\text{R}}\textbf{H}_w^\text{H}\textbf{H}_w}$ are the smallest and largest eigenvalues of the uncorrelated central complex Wishart matrix $\frac{1}{N_\text{R}}\textbf{H}_w^\text{H}\textbf{H}_w$ \cite{Telatar99ETT,Sun18TWC}, respectively, which satisfy the following properties \cite{Yin88PTRF,Bai99AP}
\begin{equation}\label{eq:ev3}
	\begin{split}
	\alpha_1^{\frac{1}{N_\text{R}}\textbf{H}_w^\text{H}\textbf{H}_w}\xrightarrow{\text{a.s.}}(1+\sqrt{\eta})^2,~\alpha_{\text{rank}(\textbf{F})}^{\frac{1}{N_\text{R}}\textbf{H}_w^\text{H}\textbf{H}_w}\xrightarrow{\text{a.s.}}(1-\sqrt{\eta})^2
	\end{split}
\end{equation}

\noindent in which "a.s." means almost sure, and $\eta=N_\text{T}/N_\text{R}$. We briefly analyze three conditions in terms of $\eta$: (1) If $N_\text{T}\ll N_\text{R}$ ($\eta\to0$), from the law of large numbers, the eigenvalues of $\textbf{F}$ in (\ref{eq:F1}) converge to those in $\tilde{\textbf{D}}_{\text{T}}$ pointwisely with high probability; (2) If $N_\text{T}\gg N_\text{R}$ ($\eta\gg1$), $\alpha_k^{\textbf{F}}\approx\eta\alpha_k^{\tilde{\textbf{D}}_{\text{T}}}$; (3) If $N_\text{T}\approx N_\text{R}$ ($\eta\approx1$), $0\lesssim\alpha_k^{\textbf{F}}\lesssim4\alpha_k^{\tilde{\textbf{D}}_{\text{T}}}$.


With the observation above at hand, we now investigate the eigenvalue characteristics of $\tilde{\textbf{B}}$ in (\ref{eq:B1}) that can be rewritten as
\begin{equation}\label{eq:B2}
\begin{split}
\tilde{\textbf{B}}=\textbf{D}_{\text{R}}\textbf{F}=N_\text{R}\tilde{\textbf{D}}_{\text{R}}\textbf{F}
\end{split}
\end{equation}

\noindent where $\tilde{\textbf{D}}_{\text{R}}=\textbf{D}_{\text{R}}/N_\text{R}$ whose eigenvalue properties have been studied in Section~\ref{sec:evR}. Since $\textbf{F}$ in (\ref{eq:F1}) is Hermitian, the condition below holds for positive eigenvalues in $\tilde{\textbf{D}}_{\text{R}}$ and non-negative eigenvalues of $\textbf{F}$ \cite[Theorem 2.2]{Xi19AMS}:
\begin{equation}\label{eq:ev4}
\begin{split}
N_\text{R}\alpha_k^{\textbf{F}}\alpha_{\text{rank}(\tilde{\textbf{D}}_{\text{R}})}^{\tilde{\textbf{D}}_{\text{R}}}\leq\alpha_k^{\tilde{\textbf{B}}}\leq N_\text{R}\alpha_k^{\textbf{F}}\alpha_1^{\tilde{\textbf{D}}_{\text{R}}}.
\end{split}
\end{equation}

\noindent Combining (\ref{eq:ev2}), (\ref{eq:ev3}), and (\ref{eq:ev4}), the following can be concluded:
\begin{equation}\label{eq:ev5}
\begin{split}
N_\text{R}\alpha_k^{\tilde{\textbf{D}}_{\text{T}}}\alpha_{\text{rank}(\tilde{\textbf{D}}_{\text{R}})}^{\tilde{\textbf{D}}_{\text{R}}}&\lesssim\alpha_k^{\tilde{\textbf{B}}}\lesssim N_\text{R}\alpha_k^{\tilde{\textbf{D}}_{\text{T}}}\alpha_1^{\tilde{\textbf{D}}_{\text{R}}},~\text{if}~N_\text{T}\ll N_\text{R},\\
N_\text{T}\alpha_k^{\tilde{\textbf{D}}_{\text{T}}}\alpha_{\text{rank}(\tilde{\textbf{D}}_{\text{R}})}^{\tilde{\textbf{D}}_{\text{R}}}&\lesssim\alpha_k^{\tilde{\textbf{B}}}\lesssim N_\text{T}\alpha_k^{\tilde{\textbf{D}}_{\text{T}}}\alpha_1^{\tilde{\textbf{D}}_{\text{R}}},~\text{if}~N_\text{T}\gg N_\text{R},\\
0&\lesssim\alpha_k^{\tilde{\textbf{B}}}\lesssim 4N_\text{R}\alpha_k^{\tilde{\textbf{D}}_{\text{T}}}\alpha_1^{\tilde{\textbf{D}}_{\text{R}}},~\text{if}~N_\text{T}\approx N_\text{R}.
\end{split}
\end{equation}

\noindent Note that the roles of $\textbf{D}_\text{T}$ and $\textbf{D}_\text{R}$ in (\ref{eq:B1}) are actually equivalent, thus (\ref{eq:F1})-(\ref{eq:ev5}) still hold if swapping the subscripts $\text{T}$ and $\text{R}$ in them. Consequently, (\ref{eq:ev5}) can be extended to (\ref{eq:ev6}). Since the lower and upper bounds of $\alpha_k^{\tilde{\textbf{B}}}$ in (\ref{eq:ev6}) are relatively loose, Monte Carlo simulations are performed to allow for more intuitive understanding. The simulation condition is $N_\text{T}=N_\text{R}$ and the RISs at the transmitter and receiver have the same dimensions, as this case renders it the most difficult to obtain tight analytical bounds on individual eigenvalues of the channel. 
\begin{figure}
	\centering
	\includegraphics[width=0.7\columnwidth]{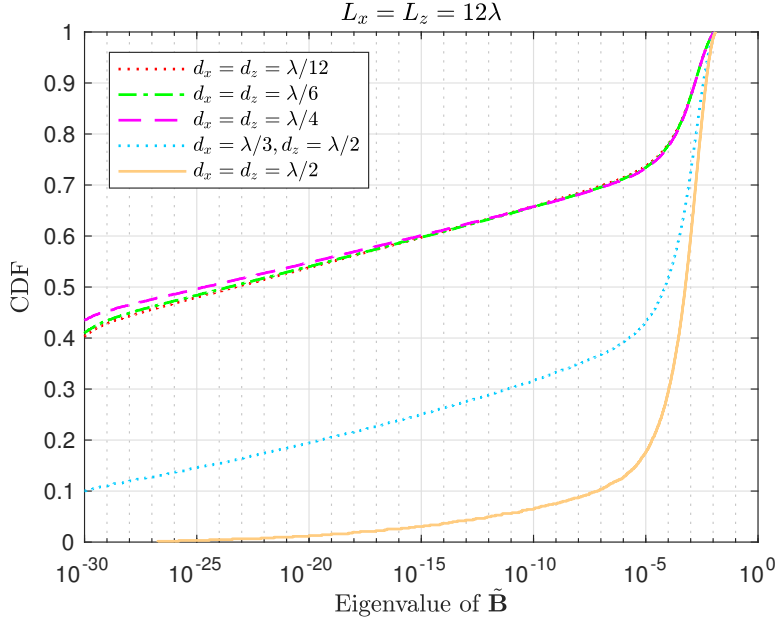}
	\caption{CDF of the eigenvalues of $\tilde{\textbf{B}}$ for various element spacing $d_x$ and $d_z$ with $L_x=L_z=12\lambda$.}
	\label{fig:fig4}	
\end{figure}

\begin{figure*}
\begin{equation}\label{eq:ev6}
\begin{split}
\max(N_\text{R}\alpha_k^{\tilde{\textbf{D}}_{\text{T}}}\alpha_{\text{rank}(\tilde{\textbf{D}}_{\text{R}})}^{\tilde{\textbf{D}}_{\text{R}}},N_\text{R}\alpha_k^{\tilde{\textbf{D}}_{\text{R}}}\alpha_{\text{rank}(\tilde{\textbf{D}}_{\text{T}})}^{\tilde{\textbf{D}}_{\text{T}}})&\lesssim\alpha_k^{\tilde{\textbf{B}}}\lesssim\min( N_\text{R}\alpha_k^{\tilde{\textbf{D}}_{\text{T}}}\alpha_1^{\tilde{\textbf{D}}_{\text{R}}},N_\text{R}\alpha_k^{\tilde{\textbf{D}}_{\text{R}}}\alpha_1^{\tilde{\textbf{D}}_{\text{T}}}),~\text{if}~N_\text{T}\ll N_\text{R},\\
\max(N_\text{T}\alpha_k^{\tilde{\textbf{D}}_{\text{T}}}\alpha_{\text{rank}(\tilde{\textbf{D}}_{\text{R}})}^{\tilde{\textbf{D}}_{\text{R}}},N_\text{T}\alpha_k^{\tilde{\textbf{D}}_{\text{R}}}\alpha_{\text{rank}(\tilde{\textbf{D}}_{\text{T}})}^{\tilde{\textbf{D}}_{\text{T}}})&\lesssim\alpha_k^{\tilde{\textbf{B}}}\lesssim\min( N_\text{T}\alpha_k^{\tilde{\textbf{D}}_{\text{T}}}\alpha_1^{\tilde{\textbf{D}}_{\text{R}}},N_\text{T}\alpha_k^{\tilde{\textbf{D}}_{\text{R}}}\alpha_1^{\tilde{\textbf{D}}_{\text{T}}}),~\text{if}~N_\text{T}\gg N_\text{R},\\
0&\lesssim\alpha_k^{\tilde{\textbf{B}}}\lesssim\min( 4N_\text{R}\alpha_k^{\tilde{\textbf{D}}_{\text{T}}}\alpha_1^{\tilde{\textbf{D}}_{\text{R}}},4N_\text{T}\alpha_k^{\tilde{\textbf{D}}_{\text{R}}}\alpha_1^{\tilde{\textbf{D}}_{\text{T}}}),~\text{if}~N_\text{T}\approx N_\text{R}.
\end{split}
\end{equation}
\end{figure*}

Under the condition of $N_\text{T}=N_\text{R}=N$, the CDF $F(\alpha^{\tilde{\textbf{B}}})$ in (\ref{eq:CDF1}) can be simplified to
\begin{equation}\label{eq:CDF2}
\begin{split}
F\Big(\alpha^{\tilde{\textbf{B}}}\Big)=\frac{1}{N}-\frac{Q_0\Big(\alpha^{\tilde{\textbf{B}}}\Big)}{N^2}\sum_{n=1}^{N}\det\bigg(\Big[\textbf{K}^{(n)}\Big]_{i,j}\bigg)
\end{split}
\end{equation}

\noindent $F(\alpha^{\tilde{\textbf{B}}})$ is illustrated in Fig.~\ref{fig:fig4} for various element spacing $d_x$ and $d_z$ which results in different values of $N$, with $L_x=L_z=12\lambda$ as an example.
\begin{figure}
	\centering
	\includegraphics[width=\columnwidth]{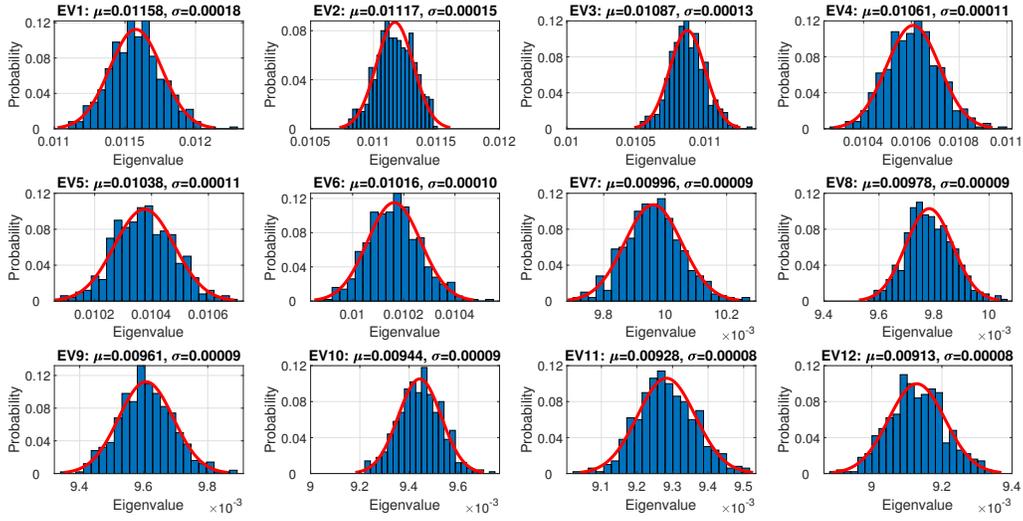}
	\caption{Probability density distributions (PDFs) of the first 12 largest eigenvalues of $\tilde{\textbf{B}}$ in (\ref{eq:B1}) with $L_x=L_z=12\lambda$ and $d_x=d_z=\lambda/12$.}
	\label{fig:fig5}	
\end{figure}
\begin{figure}
	\centering
	\includegraphics[width=0.7\columnwidth]{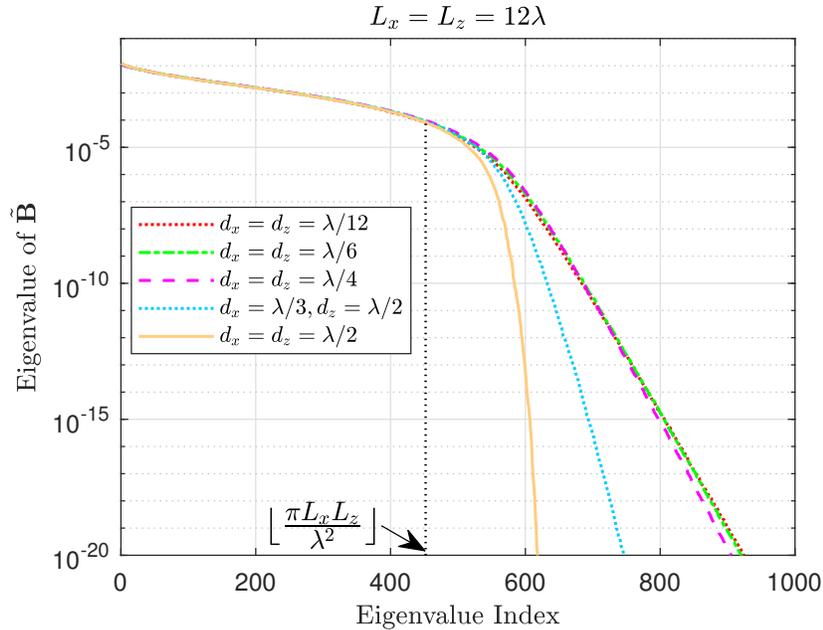}
	\caption{Eigenvalues of $\tilde{\textbf{B}}$ in (\ref{eq:B1}) in non-increasing order for various element spacing $d_x$ and $d_z$ with $L_x=L_z=12\lambda$. Also depicted is the asymptotic spatial degrees of freedom (DoF) $\lfloor\frac{\pi L_xL_z}{\lambda^2}\rfloor$ derived in \cite{Pizzo20SPAWC} for $\text{min}(L_x,L_z)/\lambda\to\infty$.}
	\label{fig:fig6}	
\end{figure}

Fig.~\ref{fig:fig5} depicts the probability density distributions (PDFs) of the first 12 largest eigenvalues of $\tilde{\textbf{B}}$ in (\ref{eq:B1}) with $L_x=L_z=12\lambda$ and $d_x=d_z=\lambda/12$, obtained from 1000 random realizations of $\textbf{H}_w$. The mean $\mu$ and standard deviation $\sigma$ via fitting using the Gaussian distribution are also displayed for each PDF. It is evident based on the tiny standard deviations that each of the eigenvalues well obey the normal distribution with a large concentration around the mean value. Fig.~\ref{fig:fig6} shows the eigenvalues of $\tilde{\textbf{B}}$ in (\ref{eq:B1}) with the same RIS dimensions and element spacing in Fig.~\ref{fig:fig3}, where each curve is averaged over 1000 random realizations of $\textbf{H}_w$. Some selected eigenvalues of $\tilde{\textbf{B}}$ are listed in Table~\ref{tbl:evB} to provide a more quantative demonstration. 

\begin{table*}
	\caption{Selected eigenvalues of $\tilde{\textbf{B}}$ in non-increasing order for various element spacing $d_x$ and $d_z$ with $L_x=L_z=12\lambda$}
	\label{tbl:evB}
	\centering
	\begin{center}
		\begin{tabular}{|p{0.03\textwidth}||p{0.13\textwidth}|p{0.13\textwidth}|p{0.13\textwidth}|p{0.13\textwidth}|p{0.13\textwidth}|}
			\hline
			& \multicolumn{5}{c|}{$\alpha_k^{\tilde{\textbf{B}}}$}\\ 
			\hline
			$k$ & $d_x=d_z=\frac{\lambda}{2}$ & $d_x=\frac{\lambda}{3},d_z=\frac{\lambda}{2}$ & $d_x=d_z=\frac{\lambda}{4}$ & $d_x=d_z=\frac{\lambda}{6}$ & $d_x=d_z=\frac{\lambda}{12}$ \\ 
			\hline\hline
			$1$ & $0.01268$ &$0.01216$&$0.01134$&$0.01148$&$0.01158$ \\ 
			\hline
			$2$ & $0.01210$ &$0.01154$&$0.01095$&$0.01107$&$0.01115$\\ 
			\hline
			$3$ & $0.01162$ &$0.01114$&$0.01066$&$0.01076$&$0.01086$ \\ 
			\hline
			$4$ & $0.01126$ &$0.01081$&$0.01040$&$0.01051$&$0.01061$ \\ 
			\hline
			$5$ & $0.01091$ &$0.01051$&$0.01018$&$0.01029$&$0.01038$\\ 
			\hline
			$6$ & $0.01062$ &$0.01025$&$0.00998$&$0.01006$&$0.01017$ \\ 
			\hline
			$7$ & $0.01038$ &$0.01003$&$0.00978$&$0.00988$&$0.00997$ \\ 
			\hline
			$8$ & $0.01013$ &$0.00980$&$0.00960$&$0.00970$&$0.00978$ \\ 
			\hline
			$9$ & $0.00991$ &$0.00961$&$0.00943$&$0.00954$&$0.00962$ \\ 
			\hline
			$10$ & $0.00971$ &$0.00942$&$0.00927$&$0.00937$&$0.00945$\\ 
			\hline
			$50$ & $0.00551$ &$0.00554$&$0.00557$&$0.00560$&$0.00561$ \\ 
			\hline
			$100$ & $0.00334$ &$0.00346$&$0.00348$&$0.00347$&$0.00348$ \\ 
			\hline
			$200$ & $0.00153$ &$0.00155$&$0.00158$&$0.00156$&$0.00155$\\ 
			\hline
			$300$ & $0.00069$ &$0.00068$&$0.00068$&$0.00065$&$0.00065$\\ 
			\hline
			$400$ & $0.00023$ &$0.00022$&$0.00023$&$0.00021$&$0.00020$\\ 
			\hline
			$500$ & $2.70\times10^{-5}$&$2.73\times10^{-5}$&$3.18\times10^{-5}$&$2.83\times10^{-5}$&$2.46\times10^{-5}$ \\ 
			\hline
			$600$ & $5.61\times10^{-14}$&$1.58\times10^{-8}$&$2.41\times10^{-7}$&$1.96\times10^{-7}$&$1.37\times10^{-7}$ \\ 
			\hline
			$700$&N/A&$2.73\times10^{-16}$&$2.46\times10^{-11}$&$2.91\times10^{-11}$&$2.23\times10^{-11}$ \\ 
			\hline
			$800$ &N/A&$7.84\times10^{-27}$&$8.27\times10^{-16}$&$1.75\times10^{-15}$&$1.77\times10^{-15}$  \\ 
			\hline
			$900$ &N/A&$0$&$1.69\times10^{-20}$&$7.83\times10^{-20}$&$1.10\times10^{-19}$ \\ 
			\hline
			$1000$ &N/A&N/A&$2.47\times10^{-25}$&$2.85\times10^{-24}$&$5.80\times10^{-24}$\\ 
			\hline
			$2000$ & N/A &N/A&$0$&$0$&$0$\\ 
			\hline
			$3000$ & N/A &N/A&N/A&$0$&$0$\\ 
			\hline
		\end{tabular}
	\end{center}
\end{table*}

\section{Effective Degrees of Freedom}
For ease of exposition, denote the sequence of the eigenvalues of $\tilde{\textbf{B}}$ in (\ref{eq:B1}) as $\{\gamma_k\}$. When $N_\text{T}\leq N_\text{R}$, and instantaneous CSI is only available at the receiver which is usually the case in practice, the channel capacity in (\ref{eq:C4}) can be recast as \cite{Shiu00TC} 
\begin{equation}\label{eq:C5} 
\begin{split}
C=\sum_{k=1}^{N_\text{S}}\mathbb{E}\Big\{\log_2\Big(1+\frac{\rho N_\text{T}N_\text{R}}{N_\text{S}}\gamma_k\Big)\Big\}.
\end{split}
\end{equation}

\noindent where $N_\text{S}$ is given in (\ref{eq:Ns}) and the text above it. To analyze the EDoF with which the capacity is maximized, an asymptotic continuous function relevant to $C$ in (\ref{eq:C5}) is defined as follows
\begin{equation}\label{eq:h1}
\begin{split}
h(N_\text{S})=\int_{1}^{N_\text{S}}\log_2\bigg(1+\frac{\rho N_\text{T}N_\text{R}}{N_\text{S}}\gamma(x)\bigg)dx
\end{split}
\end{equation}

\noindent where $h(N_\text{S})$ is a continuous function of $N_\text{S}$, and $\gamma(x)$ represents the continuous counterpart of $\{\gamma_k\}$ as illustrated in Fig.~\ref{fig:fig6}. Comparing Figs.~\ref{fig:fig4} and~\ref{fig:fig6}, it is observed that for the same $L_x$, $L_z$, and $N$ (or equivalently element spacing $d_x$ and $d_z$), the eigenvalues $\gamma(x)$ in Fig.~\ref{fig:fig6} and the CDF $F(\alpha^{\tilde{\textbf{B}}})$ in Fig.~\ref{fig:fig4} can be related as follows:
\begin{equation}\label{eq:CDF}
\begin{split}
\gamma(x)=F^{-1}\bigg(1-\frac{x}{N}\bigg)
\end{split}
\end{equation}

\noindent where $F^{-1}(\cdot)$ denotes the inverse function of the CDF $F(\cdot)$ in (\ref{eq:CDF2}). The EDoF $N_\text{S}^\star$ maximizing the capacity satisfies
\begin{equation}\label{eq:h2}
\begin{split}
\frac{dh(N_\text{S})}{dN_\text{S}}\bigg|_{N_\text{S}=N_\text{S}^\star}=0.
\end{split}
\end{equation}

\noindent Employing (\ref{eq:h1}) and the Leibniz integral rule \cite{Woods26}, $\frac{dh(N_\text{S})}{dN_\text{S}}$ is given by (\ref{eq:h3}), where the analytical form of $\gamma(x)$ is provided in (\ref{eq:CDF}). 
\begin{equation}\label{eq:h3}
\begin{split}
\frac{dh(N_\text{S})}{dN_\text{S}}=&~\log_2\bigg(1+\frac{\rho N_\text{T}N_\text{R}}{N_\text{S}}\gamma(N_\text{S})\bigg)+\frac{1}{\log2}\int_{1}^{N_\text{S}}\Bigg[\frac{1}{1+\frac{\rho N_\text{T}N_\text{R}}{N_\text{S}}\gamma(x)}\times\Bigg(-\frac{\rho N_\text{T}N_\text{R}}{N_\text{S}^2}\gamma(x)\Bigg)\Bigg]dx\\
=&~\log_2\bigg(1+\frac{\rho N_\text{T}N_\text{R}}{N_\text{S}}\gamma(N_\text{S})\bigg)-\frac{1}{N_\text{S}\log2}\int_{1}^{N_\text{S}}\frac{\frac{\rho N_\text{T}N_\text{R}}{N_\text{S}}\gamma(x)}{1+\frac{\rho N_\text{T}N_\text{R}}{N_\text{S}}\gamma(x)}dx\\
\end{split}
\end{equation}
It is straightforward from (\ref{eq:h2}) and (\ref{eq:h3}) that the EDoF $N_\text{S}^\star$ must satisfy
\begin{equation}\label{eq:EDoF}
\begin{split}
\log_2\bigg(1+\frac{\rho N_\text{T}N_\text{R}}{N_\text{S}^\star}\gamma(N_\text{S}^\star)\bigg)=\frac{1}{N_\text{S}^\star\log2}\int_{1}^{N_\text{S}^\star}\frac{\frac{\rho N_\text{T}N_\text{R}}{N_\text{S}^\star}\gamma(x)}{1+\frac{\rho N_\text{T}N_\text{R}}{N_\text{S}^\star}\gamma(x)}dx
\end{split}
\end{equation}

\noindent It is evident from (\ref{eq:EDoF}) that the EDoF $N_\text{S}^\star$ replies upon not only the eigenvalue distribution of the normalized equivalent channel $\tilde{\textbf{B}}$ in (\ref{eq:B1}), but also the SNR $\rho$ at the receiver as well as $N_\text{T}N_\text{R}$. Although analytical expressions of all the terms in (\ref{eq:EDoF}) have been obtained, it is still difficult to derive a closed-form expression of $N_\text{S}^\star$ due to the extremely complicated form of $\gamma(x)$ in (\ref{eq:CDF}). Therefore, we resort to numerical simulations to unveil the behavior of $N_\text{S}^\star$ under a variety of conditions. 
\begin{figure}
	\centering
	\includegraphics[width=0.7\columnwidth]{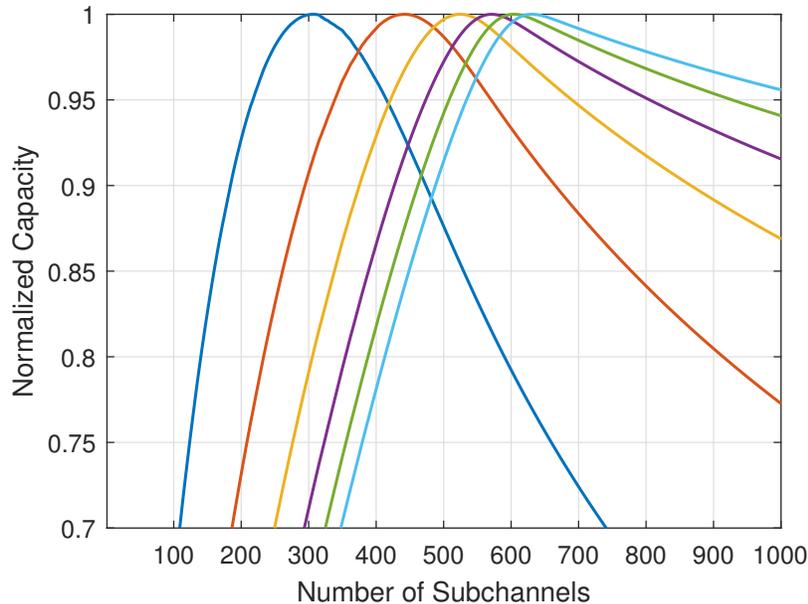}
	\caption{Normalized capacity versus number of subchannels on which data is sent, with $L_x=L_z=12\lambda$ and $d_x=d_z=\lambda/4$. The curves from leftmost to rightmost represent receive SNRs of -10 dB to 40 dB in increments of 10 dB, respectively.}
	\label{fig:fig7}	
\end{figure}
\begin{figure}
	\centering
	\includegraphics[width=0.7\columnwidth]{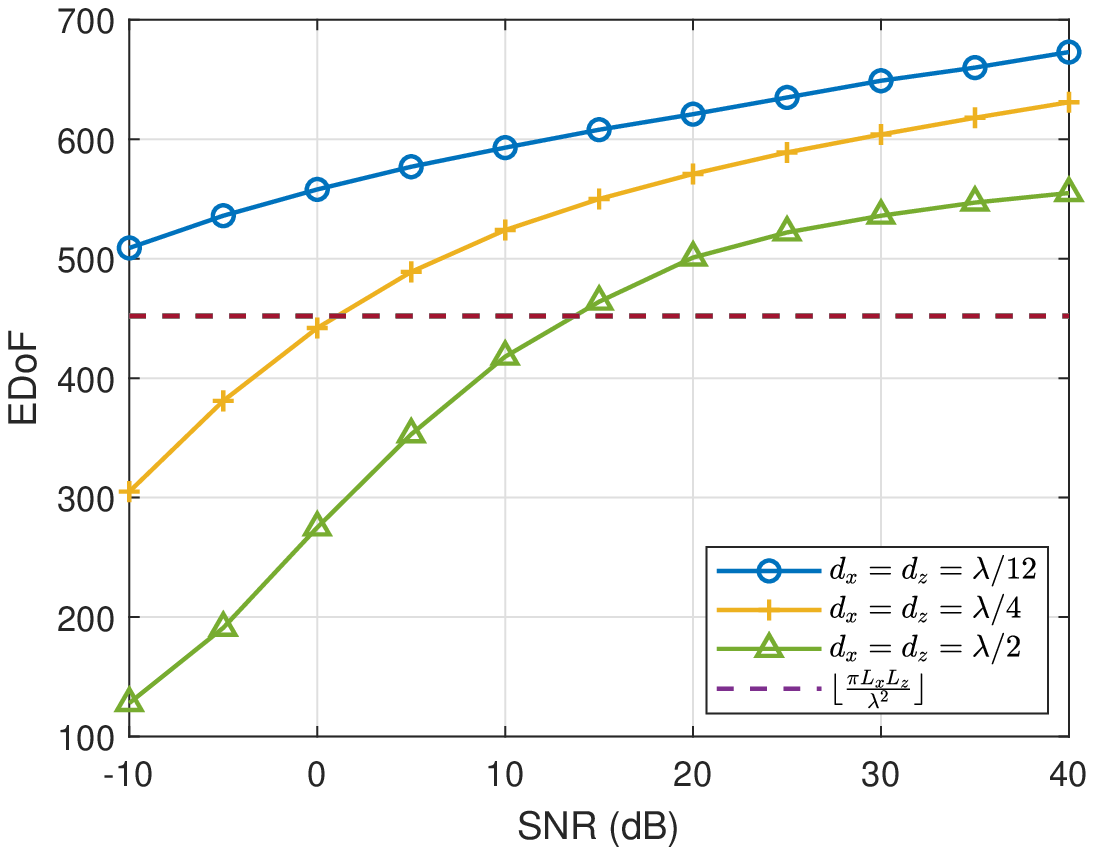}
	\caption{Effective degrees of freedom (EDoF) as a function of receive SNR for various element spacing $d_x$ and $d_z$ with $L_x=L_z=12\lambda$. The dashed straight line limns the asymptotic spatial degrees of freedom (DoF) $\lfloor\frac{\pi L_xL_z}{\lambda^2}\rfloor$ without considering SNR conditions.}
	\label{fig:fig8}	
\end{figure}
\begin{figure}
	\centering
	\includegraphics[width=0.7\columnwidth]{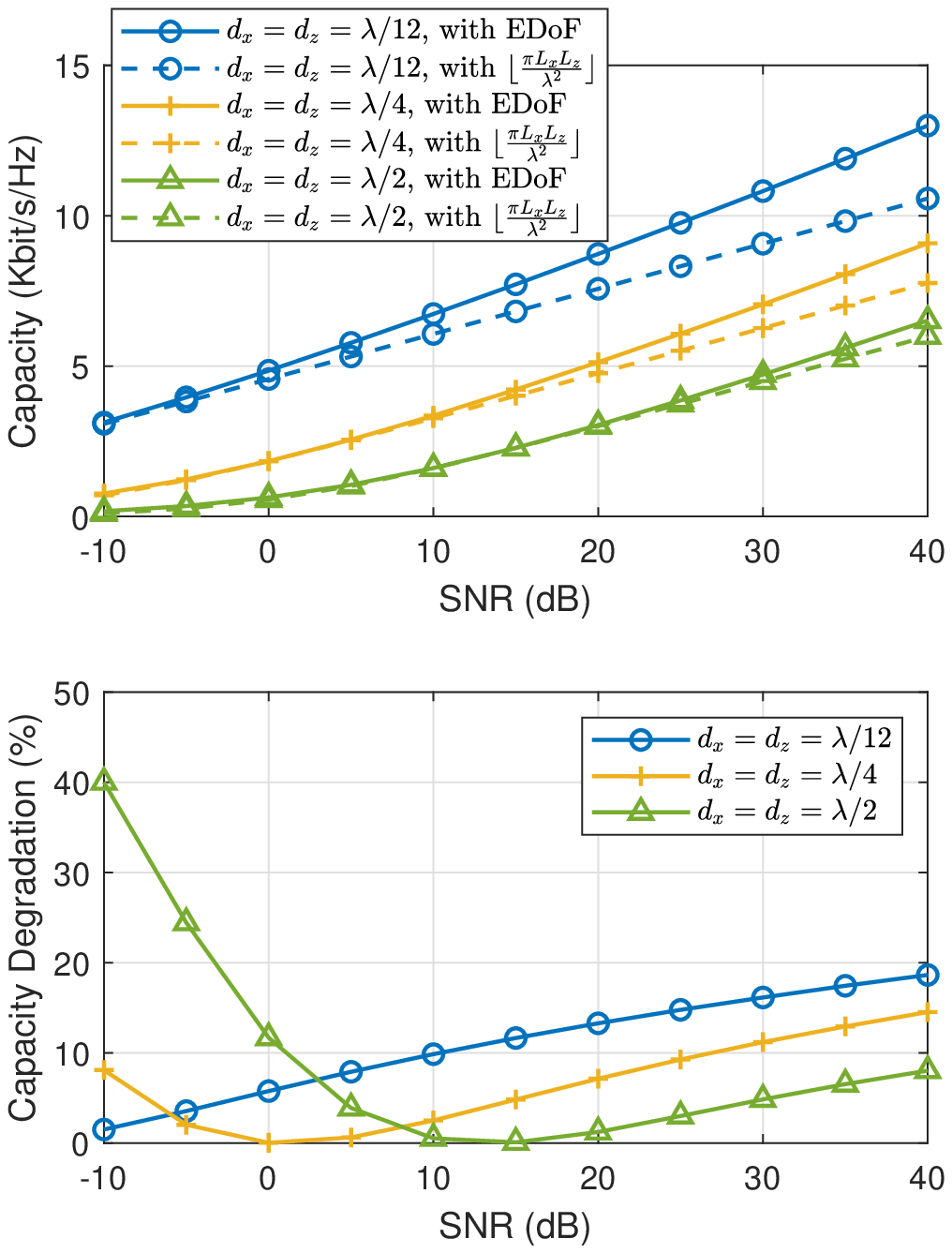}
	\caption{Upper: Capacity versus receive SNR for both EDoF and the DoF $\lfloor\frac{\pi L_xL_z}{\lambda^2}\rfloor$ for various element spacing $d_x$ and $d_z$ with $L_x=L_z=12\lambda$. Lower: Capacity degradation of using the DoF $\lfloor\frac{\pi L_xL_z}{\lambda^2}\rfloor$ against the EDoF for various element spacing $d_x$ and $d_z$ with $L_x=L_z=12\lambda$.}
	\label{fig:fig9}	
\end{figure}
\begin{figure}
	\centering
	\includegraphics[width=0.7\columnwidth]{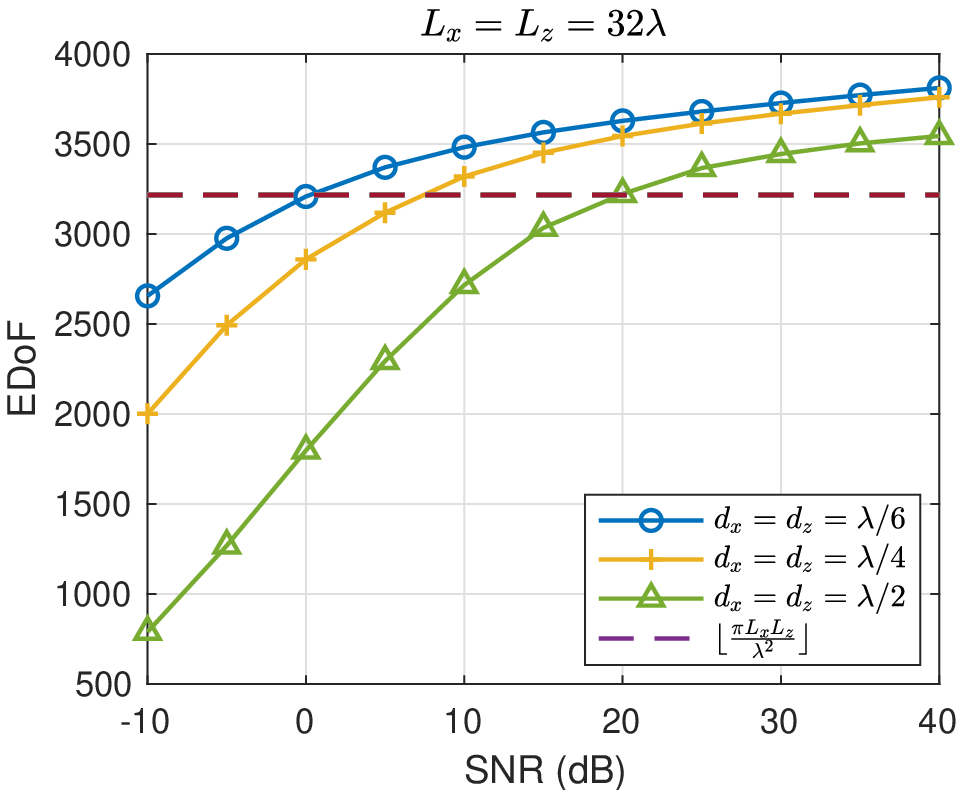}
	\caption{Effective degrees of freedom (EDoF) as a function of receive SNR for various element spacing $d_x$ and $d_z$ with $L_x=L_z=32\lambda$. The dashed straight line limns the asymptotic spatial degrees of freedom (DoF) $\lfloor\frac{\pi L_xL_z}{\lambda^2}\rfloor$ without considering SNR conditions.}
	\label{fig:fig10}	
\end{figure}
\begin{figure}
	\centering
	\includegraphics[width=0.7\columnwidth]{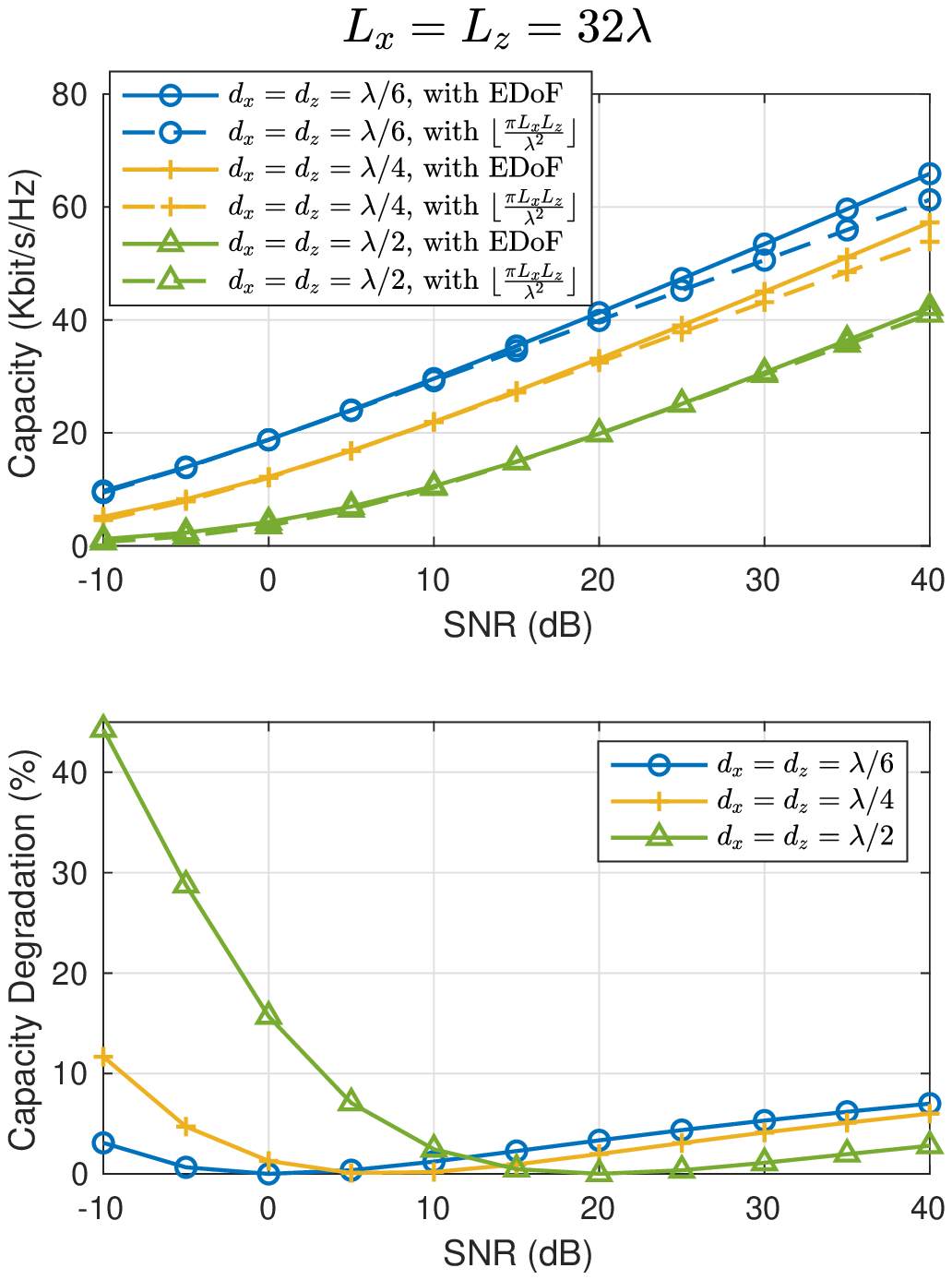}
	\caption{Upper: Capacity versus receive SNR for both EDoF and the DoF $\lfloor\frac{\pi L_xL_z}{\lambda^2}\rfloor$ for various element spacing $d_x$ and $d_z$ with $L_x=L_z=32\lambda$. Lower: Capacity degradation of using the DoF $\lfloor\frac{\pi L_xL_z}{\lambda^2}\rfloor$ against the EDoF for various element spacing $d_x$ and $d_z$ with $L_x=L_z=32\lambda$.}
	\label{fig:fig11}	
\end{figure}

Fig.~\ref{fig:fig7} displays the normalized capacity versus number of subchannels on which data is sent, with $L_x=L_z=12\lambda$ and $d_x=d_z=\lambda/4$. The curves from leftmost to rightmost represent receive SNRs of -10 dB to 40 dB in increments of 10 dB, respectively. It is obvious from Fig.~\ref{fig:fig7} that the capacity reaches its maximum value at different numbers of subchannels as the receive SNR changes, indicating that the optimal number of subchannels, i.e., the EDoF, is not fixed for given RIS configurations but varies with channel conditions, as expected and validated in practice \cite{Shiu00TC,Sun20}. To provide a more explicit comparison on the EDoF and the DoF $\lfloor\frac{\pi L_xL_z}{\lambda^2}\rfloor$ without considering the SNR condition, we show the EDoF as a function of receive SNR for various element spacing $d_x$ and $d_z$ with $L_x=L_z=12\lambda$ in Fig.~\ref{fig:fig8}, from which the following remarks can be made: First, for a given number of RIS elements $N$ or element spacing $d_x$ and $d_z$, the EDoF alters with SNR in a non-linear manner. Second, the the alteration rate and range of EDoF change with element spacing $d_x$ and $d_z$. Third, the EDoF can be smaller than, equal to, or larger than the SNR-unaware DoF $\lfloor\frac{\pi L_xL_z}{\lambda^2}\rfloor$ \cite{Pizzo20SPAWC,Sun21RISModel}, depending on the element spacing $d_x$ and $d_z$ along with SNR. This implies that the DoF $\lfloor\frac{\pi L_xL_z}{\lambda^2}\rfloor$ is not necessarily the upper bound of the EDoF. Moreover, the difference between EDoF and the SNR-unaware DoF can be huge (e.g., one can be twice of the other in some cases). 

The impact of DoF on capacity is illustrated in Fig.~\ref{fig:fig9} for various element spacing $d_x$ and $d_z$ with $L_x=L_z=12\lambda$, where the upper plot shows the capacity versus receive SNR for both EDoF and the DoF $\lfloor\frac{\pi L_xL_z}{\lambda^2}\rfloor$, and the lower plot depicts the capacity degradation of using the DoF $\lfloor\frac{\pi L_xL_z}{\lambda^2}\rfloor$ against the EDoF. As is seen, the maximum capacity degradation can reach about 40\% if sending data on the conventional $\lfloor\frac{\pi L_xL_z}{\lambda^2}\rfloor$ subchannels as opposed to the optimal number of subchannels represented by EDoF. The analysis above shows that the effective optimal number of subchannels, or equivalently spatial eigenmodes, is not solely determined by the spatial correlation of the RISs, but also by how many spatial eigenmodes are ``illuminated" by the transmit power after the weakening by large-scale fading and noise. 

To gain insights on how the EDoF and capacity behaves with different $L_x$ and $L_z$, similar simulations are performed for $L_x=L_z=32\lambda$, whose results are demonstrated in Figs.~\ref{fig:fig10} and~\ref{fig:fig11}. It is observed from that the discrepancy between EDoF and the SNR-unaware DoF $\lfloor\frac{\pi L_xL_z}{\lambda^2}\rfloor$ and the capacity degradation decrease at moderate to high SNRs, but become even more prominent at low SNRs, as compared to smaller $L_x$ and $L_z$ in Figs.~\ref{fig:fig8} and~\ref{fig:fig9}. This can be explained by the fact that as $L_x/\lambda$ and $L_z/\lambda$ increase, the accuracy of the asymptotic DoF $\lfloor\frac{\pi L_xL_z}{\lambda^2}\rfloor$ also ascends, such that the EDoF is close to $\lfloor\frac{\pi L_xL_z}{\lambda^2}\rfloor$ as long as the SNR is not too low, i.e., the spatial eigenmodes are sufficiently ``illuminated". On the other hand, the power allocated to the weak eigenmodes is very poorly used, thus the DoF error is enlarged. It is noteworthy that even for a relatively large ratio of 32 between $\min(L_x,L_z)/\lambda$, the asymptotic DoF $\lfloor\frac{\pi L_xL_z}{\lambda^2}\rfloor$ still does not serve as the upper bound of the EDoF in various settings as shown in Fig.~\ref{fig:fig10}, indicating that the actual rank of the composite channel $\tilde{\textbf{B}}$ in (\ref{eq:B1}) is larger than $\lfloor\frac{\pi L_xL_z}{\lambda^2}\rfloor$ in this case, since EDoF should lie between 0 and the channel rank \cite{Shiu00TC}. 

\section{Conclusion}
In this paper, we have systematically investigated the characteristics of eigenvalues of the spatial correlation matrix at the RIS and those of the composite channel matrix with RISs at both the transmitter and receiver. More importantly, we have also studied the EDoF of the composite channel that yields the maximum capacity. Distinct from the DoF of RISs in the existing literature which ignores the SNR condition, EDoF is affected by SNR as well as the spatial correlation. Although the analysis in this paper is conducted for the isotropic scattering environment, it is expected that similar trends on the EDoF will occur for non-isotropic scattering, with fewer EDoF in general due to more severe spatial correlation \cite{Sun21RISModelArxiv}.

\ifCLASSOPTIONcaptionsoff
  \newpage
\fi

\bibliographystyle{IEEEtran}
\bibliography{RIS}

\end{document}